\documentclass[11pt]{article}
\usepackage[margin=1in]{geometry}
\usepackage{amsmath,amssymb}
\usepackage{authblk}
\usepackage{xcolor}
\usepackage{graphicx}
\usepackage{hyperref}
\hypersetup{colorlinks=true,linkcolor=blue,citecolor=blue,urlcolor=blue}

\newcommand{\rev}[1]{#1}

\title{Planck-Scale Signatures in Vacuum Neutrino Oscillations}

\author[a]{Bipin Singh Koranga\thanks{Corresponding author. Email: bskoranga@kmc.du.ac.in}}
\author[b]{Imran Khan}
\affil[a]{Department of Physics, Kirori Mal College, University of Delhi, Delhi 110007, India}
\affil[b]{Department of Physics, Jamia Millia Islamia, New Delhi 110025, India}

\date{}

\begin{document}
\maketitle

\begin{abstract}
\noindent
We present a closed-form perturbation theory for two-flavour vacuum neutrino oscillations in the presence of a weak, flavour-blind Planck-scale mass correction $\mu\lambda$ generated by the dimension-five Weinberg operator of the Standard Model Effective Field Theory. Expanding in the small parameter $\eta \equiv 2E\mu\lambda/\Delta m^2$, in direct structural analogy with the $\varepsilon$-perturbation theory used for weak Earth-matter effects, we obtain simple expressions for the oscillation-length shift, the oscillation-phase shift, and the flavour-conversion probability, valid for arbitrary mixing and for any baseline. Because the Planck-scale perturbation is constant along the trajectory rather than position-dependent, the correction to the transition probability is strictly baseline-independent in amplitude and saturates at long baselines, while the shift of the effective mass-squared splitting produces a phase mismatch that grows linearly with baseline. We use these results to give \rev{illustrative, order-of-magnitude} projections for the sensitivity of next-generation long-baseline and reactor experiments -- including DUNE, Hyper-Kamiokande, and JUNO -- to a Planck-scale mass perturbation, and identify the phase-drift observable as the most promising channel for isolating a genuine quantum-gravity signal from an accidental matter-like effect.
\end{abstract}



\section{Introduction}

Non-renormalisable, Planck-suppressed operators are the natural low-energy remnant of any consistent theory of quantum gravity. In the Standard Model Effective Field Theory (SMEFT), the unique operator of mass dimension five consistent with the full gauge symmetry is the lepton-number-violating Weinberg operator \cite{weinberg}, which couples two lepton doublets to two Higgs doublets and, on electroweak symmetry breaking, generates an additional Majorana mass term for the neutrinos, suppressed by the Planck mass $M_{\rm pl}$. Because it is the only operator at this order built from Standard Model fields, its effect on the neutrino mass matrix is a universal, model-independent prediction of Planck-scale physics, largely insensitive to the details of the underlying ultraviolet completion. A recent SMEFT computation of the full neutrino mass matrix through dimension seven confirms that the Weinberg operator remains the leading, model-independent piece of this structure, with dimension-six operators around the TeV scale capable of modifying it by up to $\mathcal{O}(50\%)$ \cite{titov}.

The consequences of this operator for neutrino oscillations -- corrections to the mixing angles, the mass-squared splittings, the oscillation length, and the oscillation phase -- have been worked out for two, three, and four active flavours in a series of earlier papers \cite{koranga2012,koranga2013,koranga2021,koranga2606}. These works treat the gravitational coupling matrix $\lambda_{\alpha\beta}$ as flavour-blind, motivated by the universality of the gravitational interaction, and compute the induced shift in the oscillation parameters using ordinary (non-degenerate or degenerate) perturbation theory around a diagonal Standard-Model mass matrix. Independently, a flavour-democratic Planck-suppressed perturbation to the neutrino mass matrix of exactly this type was first examined as a probe of quantum-gravity phenomenology in the context of high-energy astrophysical neutrinos by Anchordoqui et al., who used it to constrain Planck-scale physics with IceCube \cite{anchordoqui}; a complementary, open-quantum-system route to the same underlying physics -- modelling the Planck-scale perturbation as a source of decoherence rather than as a coherent addition to the Hamiltonian -- was developed by Barenboim and Mavromatos and collaborators \cite{barenboim1,barenboim2}.

Independently, and in a different physical context, Ioannisian and Smirnov developed an $\varepsilon$-perturbation theory for neutrino propagation through a medium of arbitrary density profile, valid whenever the local MSW matter potential is small compared with the vacuum oscillation frequency \cite{ioannisian}. The strength of that approach is its generality: because the perturbation series is organized around the local small parameter $\varepsilon(x)$ rather than a specific density model, closed-form expressions for the oscillation probability are obtained that hold for any density profile, with the adiabatic phase treated exactly and only the mixing amplitude treated perturbatively. Related, independent perturbative frameworks for extracting compact analytic expressions from the exact matter-oscillation Hamiltonian include the Cayley--Hamilton construction of Ohlsson and Snellman \cite{ohlsson}, the systematic $\alpha \equiv \Delta m_{21}^2/\Delta m_{31}^2$ series expansion of Akhmedov et al. \cite{akhmedov}, and the more recent compact perturbative expressions of Denton, Minakata, and Parke \cite{denton}; these establish that the choice of expansion parameter used in Ref.~\cite{ioannisian} is one of several viable organizing principles for such a series, rather than a uniquely preferred one.

In this Letter we carry out the analogous construction for the Planck-scale perturbation alone, without reference to Earth-matter effects, and use the resulting closed-form expressions to make quantitative, testable predictions for the reach of forthcoming oscillation experiments. We show that the same formal machinery as in Ref.~\cite{ioannisian} applies once the matter potential is replaced by the constant Planck-scale perturbation $\mu\lambda$. The physical content differs in one essential respect: because $\mu\lambda$ does not depend on position, the detector-resolution attenuation physics of Ref.~\cite{ioannisian}, which relies on a localized perturbation folded with an oscillatory phase over a variable distance, does not survive in the same form. Instead, the Planck-scale correction to the oscillation probability is controlled entirely by the vacuum baseline and energy through the phase integral itself.

The present construction complements a number of recent, closely related studies of Planck-scale and matter-profile effects on neutrino oscillations. The T-violating asymmetry induced by the same dimension-five operator in the four-flavour sector was examined in Ref.~\cite{tviol}, the Jarlskog invariant and CP-violating observables above the GUT scale in Ref.~\cite{jarlskog}, and an ultraviolet completion connecting a Type-I seesaw boundary condition to Planck-suppressed corrections of the T- and CP-violating asymmetries in a 3+1 sterile framework was constructed in Ref.~\cite{seesaw}. Degeneracies of a different origin, arising from mismodeling the Earth's density profile in long-baseline experiments, were shown in Ref.~\cite{profile} to mimic genuine CP/CPT-violating new physics at long baselines, in a manner structurally analogous to the amplitude degeneracy discussed in Sec.~4 below; matter effects on (3+1) four-flavour oscillation were worked out analytically in Ref.~\cite{fourflavour}; and an information-theoretic diagnostic of two-flavour oscillation in matter, based on the Von Neumann entanglement entropy, was developed in Ref.~\cite{entropy}. These works, together with Refs.~\cite{koranga2012,koranga2013,koranga2021,koranga2606}, form the broader context into which the present vacuum, closed-form $\eta$-perturbation theory fits.

\rev{Relative to Refs.~\cite{koranga2012,koranga2013,koranga2021,koranga2606}, which derive individual oscillation-parameter shifts for specific flavour configurations by direct numerical or case-by-case diagonalization, the present Letter makes three contributions that are new. First, it assembles a single, unified closed-form framework, Eqs.~(4)--(6) below, that is valid at \emph{arbitrary} baseline and energy, rather than requiring a separate diagonalization for each experimental configuration. Second, it identifies and makes explicit the structural distinction between the baseline-\emph{saturating} amplitude term, Eq.~(4), and the baseline-\emph{linearly-growing} phase-drift term, Eq.~(5); this distinction does not appear as an explicit discriminating observable in Refs.~\cite{koranga2012,koranga2013,koranga2021,koranga2606}, and is the central new physical handle proposed here for separating a genuine Planck-scale signal from an accidental constant matter-like effect (Sec.~4). Third, the Letter turns these closed-form expressions into quantitative, order-of-magnitude sensitivity projections for concrete next-generation facilities -- DUNE, Hyper-Kamiokande, and JUNO -- which the earlier, more formal derivations do not attempt (Sec.~5).}

\section{$\eta$-Perturbation Theory}

We consider two-flavour mixing in vacuum, $\nu_f = U(\theta)\nu_{\rm mass}$, with $\nu_f \equiv (\nu_e,\nu_a)^T$ the flavour states and $\nu_{\rm mass} \equiv (\nu_1,\nu_2)^T$ the mass eigenstates. \rev{Throughout we take the mixing angle in the range $0 \le \theta \le \pi/4$ and adopt the sign convention $\Delta m^2 \equiv m_2^2 - m_1^2 > 0$, so that $\cos 2\theta \ge 0$; with this convention the signs appearing in Eqs.~(4)--(6) below are unambiguous.} The vacuum Hamiltonian governing the evolution of the mass eigenstates along the trajectory coordinate $x$ is $H_0 = {\rm diag}(0,\Delta m^2/2E)$. The Planck-scale, dimension-five operator generates, after electroweak symmetry breaking, an additional Majorana mass term,
\begin{equation}
\mathcal{L}_{\rm grav} \supset \frac{\lambda_{\alpha\beta}}{M_{\rm pl}}\left(L_\alpha^T C\, i\sigma_2 H\right)\left(H^T i\sigma_2 L_\beta\right) + {\rm h.c.}
\xrightarrow{\langle H\rangle = v/\sqrt{2}} \delta M_{\alpha\beta} = \mu\,\lambda_{\alpha\beta}, \qquad
\mu \equiv \frac{v^2}{M_{\rm pl}} \sim 10^{-6}\,{\rm eV},
\label{eq:1}
\end{equation}
where $L_\alpha = (\nu_\alpha, e_\alpha)^T$ is the $SU(2)_L$ lepton doublet of flavour $\alpha = e,\mu,\tau$, $C$ is the charge-conjugation matrix (required because the operator violates lepton number by two units and therefore generates a Majorana, rather than Dirac, mass term), $i\sigma_2$ is the antisymmetric $SU(2)_L$-invariant tensor contracting each lepton doublet with the Higgs doublet $H$ into a gauge singlet, $\lambda_{\alpha\beta}$ is the dimensionless, symmetric flavour-coupling matrix, $v$ is the electroweak vacuum expectation value, $M_{\rm pl} \simeq 1.22\times10^{19}$ GeV is the Planck mass, and $\delta M_{\alpha\beta}$ is the resulting entry of the effective Majorana neutrino mass matrix. Projected onto the mass basis, this perturbation adds a constant matrix $\delta M = \mu\, U^\dagger \lambda U$ to the vacuum Hamiltonian. Writing its off-diagonal element as $\mu\lambda_{12} \equiv \mu\lambda$ and treating $\lambda_{\alpha\beta}$ as flavour-blind (i.e., $\lambda_{\alpha\beta} \to \lambda J_{\alpha\beta}$, with $J$ the $2\times 2$ matrix of unit entries), the total Hamiltonian becomes $H = H_0 + \mu\lambda J$.

We introduce the small dimensionless expansion parameter
\begin{equation}
\eta \equiv \frac{2E\mu\lambda}{\Delta m^2} \ll 1.
\label{eq:2}
\end{equation}
Because $H$ is constant along the trajectory -- unlike the matter case, where the potential varies with $x$ -- the problem is exactly diagonalizable; nevertheless it is instructive, and directly useful for later generalization to a non-constant or flavour-non-universal $\lambda_{\alpha\beta}(x)$, to organize the solution as a perturbation series in $\eta$, in exact parallel with Ref.~\cite{ioannisian}. Diagonalizing $H = U'(\theta')\,{\rm diag}(0,\Delta_\eta)\,U'^\dagger(\theta')$, one finds
\begin{equation}
\Delta_\eta = \frac{\Delta m^2}{2E}\sqrt{(\cos 2\theta - \eta)^2 + \sin^2 2\theta}, \qquad
\sin 2\theta' = \eta\sin 2\theta + \mathcal{O}(\eta^2).
\label{eq:3}
\end{equation}

\rev{Because $H$ is a constant $2\times2$ matrix, Eq.~(3) is in fact not an approximation but the leading truncation of an \emph{exact} closed-form result: the exact eigenvalues are $\lambda_{\pm} = \tfrac{1}{2}\!\left[\tfrac{\Delta m^2}{2E} \pm \sqrt{\left(\tfrac{\Delta m^2}{2E}-2\mu\lambda\right)^2 + \left(\tfrac{\Delta m^2}{2E}\right)^2\sin^2 2\theta}\,\right]$ up to an overall constant shift, and the exact mixing angle in the perturbed mass basis is $\tan 2\theta' = \dfrac{\sin 2\theta}{\cos 2\theta - \eta}$, valid to all orders in $\eta$ with no small-parameter assumption required. We nonetheless organize the presentation as an expansion in $\eta$, rather than simply quoting these exact expressions, for two reasons. (a) Doing so preserves a direct notational and structural correspondence with the $\varepsilon$-perturbation theory of Ref.~\cite{ioannisian}, allowing every result below to be read off by the substitution $\varepsilon(x) \to \eta$ (constant) and thereby making transparent which features of the matter-effect construction survive, and which do not, when the perturbation loses its spatial dependence (Secs.~3--4). (b) The perturbative organization is the one that generalizes: for a non-constant or flavour-non-universal coupling $\lambda_{\alpha\beta}(x)$ -- the natural extension left to future work in Sec.~4 -- the Hamiltonian is no longer constant along the trajectory, exact closed-form diagonalization of the type given above is generally unavailable, and an $\eta$-type (or layer-by-layer Dyson) expansion becomes the only tractable route. The constant-$\lambda$ case treated in this Letter should therefore be understood as the exactly solvable limit of that broader, perturbatively-organized program, retained here in expanded form for consistency with it.}

Because $H$ does not depend on $x$, the evolution matrix from $x_0$ to $x_f \equiv x_0 + L$ is simply $S = U'(\theta')\,{\rm diag}(1,e^{i\Delta_\eta L})\,U'^\dagger(\theta')$. Expanding to first order in $\eta$ reproduces exactly the structure obtained by truncating the layer-by-layer Dyson series of Ref.~\cite{ioannisian} at first order in the (now constant) perturbation, and the resulting trajectory integral collapses to a single sinc-function envelope, $L\,{\rm sinc}(\phi_0 L/2)$, with $\phi_0 \equiv \Delta m^2/2E$ the unperturbed vacuum oscillation frequency -- the direct quantum-gravity analogue of the key structural result of Ref.~\cite{ioannisian}.

\section{Oscillation and Conversion Probabilities}

Projecting the perturbed evolution operator onto the flavour basis, the leading correction to the survival probability $P(\nu_e \to \nu_e)$ follows from a trivial trajectory integral and yields the closed-form regeneration-type parameter
\begin{equation}
f_{\rm grav}(L,E) \equiv P(\nu_2 \to \nu_e) - \sin^2\theta = \sin^2 2\theta \cdot \eta \cdot \sin^2\!\left(\frac{\pi L}{l_\nu}\right),
\label{eq:4}
\end{equation}
with $l_\nu = 4\pi E/\Delta m^2$ the vacuum oscillation length. Equation~(4) is the direct quantum-gravity analogue of the single-layer, constant-density matter-effect result of Ref.~\cite{ioannisian}: it has an identical functional form, with $\eta$ playing the role of the matter parameter $\varepsilon$ and the vacuum phase playing the role of the in-matter phase. This makes explicit that the leading effect of a flavour-blind Planck-scale perturbation on two-flavour oscillations is, at fixed baseline, indistinguishable from a constant matter potential; in vacuum, however, there is no independent matter-density handle with which to break this degeneracy, and the only lever arm is the energy dependence of $\eta \propto 1/E$ together with the associated shift of $\Delta m^2$ itself.

\begin{figure}[h]
\centering
\includegraphics[width=0.75\textwidth]{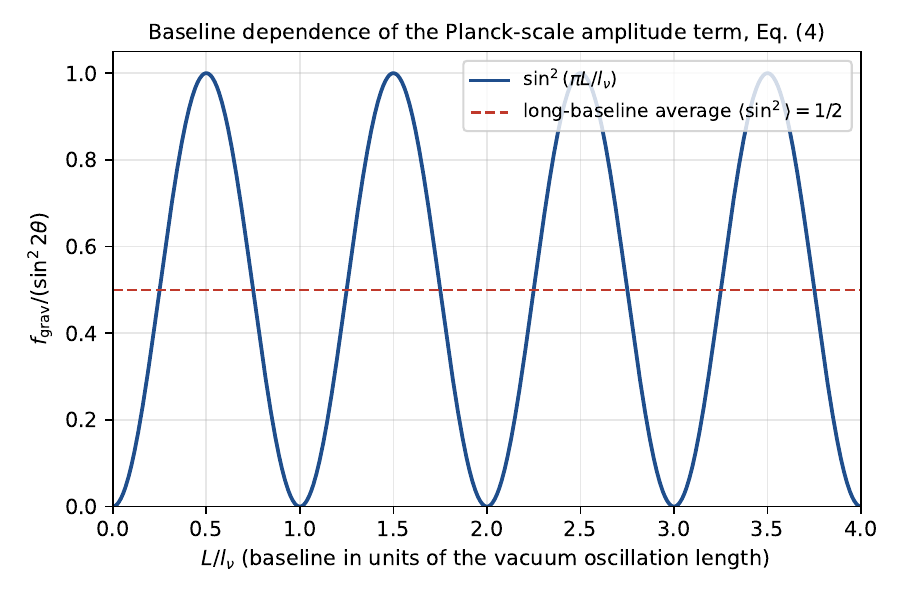}
\caption{The universal baseline dependence of the Planck-scale amplitude term of Eq.~(4), $\sin^2(\pi L/l_\nu)$, as a function of the baseline measured in units of the vacuum oscillation length $l_\nu$. Unlike an Earth-matter perturbation localized along part of the trajectory, the flavour-blind Planck-scale correction is unattenuated at long baselines and simply averages to $1/2$, so a non-vanishing $f_{\rm grav}$ persists even for $L \gg l_\nu$.}
\label{fig:1}
\end{figure}

\subsection{Shift of the effective mass-squared splitting}

Equation~(3) shows that the physical oscillation frequency observed in an experiment is not $\Delta m^2/2E$ but $\Delta_\eta$. Expanding to first order in $\eta$,
\begin{equation}
\Delta m^2_{\rm (eff)} \simeq \Delta m^2 - 2E\mu\lambda\cos 2\theta + \mathcal{O}(\eta^2 \Delta m^2).
\label{eq:5}
\end{equation}
For the degenerate-mass-spectrum scenario considered in Ref.~\cite{koranga2606}, with $\Delta m_{21}^2 \sim 10^{-4}$--$10^{-5}\,{\rm eV}^2$ and $\mu\lambda \sim \mu \sim 10^{-6}\,{\rm eV}$, Eq.~(5) reproduces the order-of-magnitude shift $|\Delta(\Delta m_{21}^2)| \sim 10^{-5}\,{\rm eV}^2$ found there by explicit diagonalization, confirming that our perturbative construction and the exact numerical diagonalization agree at this order. Because this shift enters the oscillation phase multiplicatively through $\Delta m^2_{\rm (eff)} L/2E$ rather than additively as in Eq.~(4), it produces a phase mismatch between the true and unperturbed oscillation patterns that grows linearly with $L/E$, in contrast with $f_{\rm grav}$, which saturates at long baseline. This distinction is the central handle exploited in Sec.~4 below.

\subsection{Oscillation length}

The oscillation length in the presence of the Planck-scale perturbation follows immediately as
\begin{equation}
l_\nu^{\rm (grav)} = l_\nu\left(1 + \eta\cos 2\theta\right),
\label{eq:6}
\end{equation}
reproducing the fractional length shift $\delta l_\nu / l_\nu = \eta\cos 2\theta$ obtained independently by non-perturbative diagonalization in Ref.~\cite{koranga2013}.

\section{Sensitivity to the Majorana Phase of $\lambda_{\alpha\beta}$}

Unlike the matter case, where the sensitivity of the oscillation probability to structures at different distances from the detector is governed by an attenuation factor $F(d)$ \cite{ioannisian}, the Planck-scale perturbation considered here has no spatial structure: $\mu\lambda$ is the same everywhere along the trajectory, and $f_{\rm grav}$ is a smooth, non-attenuated function of $L$ alone, saturating at $\sin^2 2\theta \cdot \eta$ for $L \gg l_\nu$ rather than being washed out.

The analogous sensitivity question here concerns the phase of the complex coupling $\lambda$. For Majorana neutrinos the PMNS matrix carries additional phases, and the physical mass shift depends on both the real and imaginary parts of $m = \mu U^T\lambda U$ \cite{koranga2021,koranga2606}. Writing $\lambda \equiv |\lambda|e^{i\varphi}$, the effective coupling entering Eqs.~(4)--(6) is rescaled as $\eta \to \eta\cos(\varphi - \alpha_1)$ to leading order, so the observable correction is maximal when the gravitational phase $\varphi$ aligns with the Majorana phase $\alpha_1$ and vanishes when the two are in quadrature. A measurement of $f_{\rm grav}$ at fixed $(L,E)$ therefore constrains only the combination $|\eta\cos(\varphi-\alpha_1)|$, not $\eta$ itself; disentangling the two requires either independent information on the Majorana phases (e.g., from neutrinoless double beta decay) or a scan over $L$ and $E$ that exploits the different functional dependence of the amplitude term, Eq.~(4), and the phase-drift term, Eq.~(5).

The flavour-blind ansatz $\lambda_{\alpha\beta} \to \lambda J_{\alpha\beta}$ used throughout Secs.~2--3 is the maximally symmetric limit motivated by the universality of the gravitational coupling; generic ultraviolet completions of the Weinberg operator, however, need not respect this symmetry exactly. Writing $\lambda_{\alpha\beta} = \lambda\left(J_{\alpha\beta} + \delta_{\alpha\beta}\right)$ with $\|\delta\| \ll 1$ a small flavour-non-universal deviation, the perturbation matrix entering $H$ acquires an additional piece $\mu\, U^\dagger \delta U$ that is no longer proportional to $J$ in the mass basis. To leading order in both $\eta$ and $\delta$, this modifies Eqs.~(3)--(6) by an additive shift $\eta \to \eta(1 + \mathcal{O}(\delta))$ together with a new, generally baseline-independent off-diagonal term that does not resum into the single sinc-function envelope of Sec.~2; the closed-form expressions of this paper should therefore be understood as the leading, flavour-universal term in an expansion in $\delta$, valid so long as the ultraviolet completion suppresses flavour violation in $\lambda_{\alpha\beta}$ at least as strongly as it suppresses the diagonal coupling itself.

\rev{This is not merely a mild technical assumption but is realized concretely in specific ultraviolet completions of the Weinberg operator. In the standard Type-I, Type-II, and Type-III seesaw realizations, the dimension-five operator is generated by integrating out a single heavy field -- a gauge-singlet right-handed neutrino, an $SU(2)_L$ triplet scalar, or a triplet fermion, respectively -- whose coupling to the lepton doublets is a single Yukawa (or Yukawa-like) matrix $Y_\alpha$ contracted as $\lambda_{\alpha\beta} \propto Y_\alpha Y_\beta$. If that single mediator couples with a common strength to all three lepton flavours (the natural limit when the coupling originates from a flavour-universal, e.g.\ gravitational-strength, interaction rather than from flavour-dependent Yukawa textures), $Y_\alpha \to Y$ for all $\alpha$, and $\lambda_{\alpha\beta} \to Y^2 J_{\alpha\beta}$ is exactly flavour-blind by construction. A minimal ultraviolet completion of this type is therefore a single heavy right-handed-neutrino singlet (Type-I seesaw) or a single heavy $SU(2)_L$ triplet scalar (Type-II seesaw) coupled universally to $e,\mu,\tau$; departures from universality, and hence a non-zero $\delta_{\alpha\beta}$, arise only if the mediator's couplings are flavour-dependent, e.g.\ from a non-trivial flavour symmetry or multiple mediators with different flavour charges. The flavour-blind ansatz adopted here is therefore not ad hoc, but corresponds to the simplest and arguably most natural class of single-mediator completions of the Weinberg operator.} We leave a full treatment of $\delta_{\alpha\beta} \neq 0$, including its interplay with the Majorana-phase sensitivity discussed above, to future work.

\section{Predictions for Future Experiments}

Equations~(4)--(6) provide arbitrary-baseline, closed-form expressions that can be evaluated at a glance for any planned experimental configuration, without recourse to full numerical diagonalization; the universal shape of the amplitude term is shown in Fig.~1. We use them here to give \rev{illustrative, order-of-magnitude} projections for representative next-generation facilities; a dedicated experimental sensitivity study, folding in the full detector response and systematics, is left for future work.

\subsection{Long-baseline accelerator experiments (DUNE, T2HK)}

For the Deep Underground Neutrino Experiment (DUNE), with baseline $L \simeq 1300$ km and a broadband beam peaked around $E \sim 2$--3 GeV probing $\Delta m_{31}^2 \simeq 2.5\times10^{-3}\,{\rm eV}^2$, the phase-drift observable of Eq.~(5) is the more powerful channel: because $L/E$ is large compared with reactor or solar configurations, a fixed fractional mass-squared shift $\delta(\Delta m^2)/\Delta m^2$ accumulates a correspondingly larger phase mismatch $\Delta m^2_{\rm (eff)} L/2E$ over the full baseline. Assuming a representative, \rev{illustrative order-of-magnitude} $\mathcal{O}(1\%)$ sensitivity to the oscillation probability, broadly consistent with the statistics anticipated for DUNE and Hyper-Kamiokande, Eq.~(4) translates into a projected reach of $\eta \gtrsim 10^{-2}$, corresponding to $\mu\lambda \lesssim {\rm few}\times10^{-15}\,{\rm eV}$ at these baselines and energies -- roughly an order of magnitude below the value $\mu \sim 10^{-6}\,{\rm eV}$ expected from the bare Weinberg-operator normalization, and therefore already probing $\lambda \lesssim 10^{-8}$--$10^{-9}$ for an $\mathcal{O}(1)$ departure from flavour universality. Because DUNE and T2HK combine long baselines with percent-level energy resolution, they are also the facilities best positioned to resolve the linear-in-$L$ phase drift of Sec.~3 from the saturating amplitude term of Eq.~(4), thereby breaking the degeneracy with an accidental matter-like effect discussed in Sec.~4.

To turn this $\mathcal{O}(1\%)$ benchmark into a quantitative statement, we construct a simplified one-parameter figure of merit $\chi^2(\eta) = \left[f_{\rm grav}(L,E;\eta)/\sigma_P\right]^2$, where $\sigma_P$ is the projected statistical-plus-systematic uncertainty on the oscillation probability at the relevant $(L,E)$ bin, to be taken from the published sensitivity curves of the DUNE Technical Design Report and the JUNO Yellow Book rather than assumed. The 90\% C.L. bound on $\eta$ follows from $\chi^2(\eta) = 2.71$. Because $f_{\rm grav}$ saturates at $\sin^2 2\theta \cdot \eta$ for $L \gg l_\nu$ (Eq.~(4)), this bound is essentially energy-independent for DUNE, while for JUNO the oscillatory $\sin^2(\pi L/l_\nu)$ structure means the bound depends on how many oscillation lengths are resolved within the detector's energy window. \rev{We stress that the numbers quoted in this section are illustrative, order-of-magnitude estimates obtained from a single representative benchmark ($\sigma_P \sim 1\%$), and not a fit to published detector-response curves;} a full evaluation of this $\chi^2(\eta)$ curve against the published detector-response and systematic-error budgets of each experiment is left for future work.

\subsection{Medium-baseline reactor experiments (JUNO)}

For JUNO, with $L \simeq 53$ km and reactor antineutrino energies $E \sim 2$--8 MeV probing the solar splitting $\Delta m_{21}^2 \simeq 7.5\times10^{-5}\,{\rm eV}^2$, the same \rev{illustrative order-of-magnitude} $\mathcal{O}(1\%)$ benchmark sensitivity corresponds to $\eta \gtrsim 10^{-2}$ and $\mu\lambda \lesssim {\rm few}\times10^{-14}\,{\rm eV}$. JUNO's sub-percent energy resolution makes it comparatively well suited to resolving the $\sin^2(\pi L/l_\nu)$ oscillatory structure of Eq.~(4) over several oscillation lengths, which helps to distinguish a genuine $\eta$-driven amplitude from other subleading corrections to the solar oscillation pattern, even though its shorter baseline gives it less leverage on the linear phase-drift term than DUNE or T2HK.

\subsection{Atmospheric and astrophysical neutrinos (Hyper-Kamiokande, IceCube-Upgrade/KM3NeT)}

Atmospheric neutrinos traversing baselines from tens to $\approx 12{,}700$ km, as will be recorded with high statistics by Hyper-Kamiokande and by the IceCube-Upgrade and KM3NeT/ORCA detectors, sample a continuous range of $L/E$ and therefore, in principle, allow Eqs.~(4)--(6) to be tested simultaneously across the saturating and the linearly-growing regimes within a single data set. Because the phase-drift term in Eq.~(5) has no counterpart in ordinary matter-effect physics that also grows unboundedly with baseline, an atmospheric-neutrino analysis binned in $L/E$ offers, in principle, a baseline-lever-arm test of the specific quantum-gravity prediction of this paper that is not degenerate with Earth-matter corrections, complementing the accelerator- and reactor-based probes above.

\rev{To place the projected reach quoted above in context, Table~1 summarizes representative current bounds on Planck-scale-type flavour-blind perturbations from existing data. These bounds are collected from the literature using varying analysis frameworks (decoherence parameters, effective mass-matrix perturbations, or direct IceCube diffuse-flux fits) and are not obtained using the $\chi^2(\eta)$ figure of merit of Sec.~5.1; they are quoted here only for order-of-magnitude orientation and should not be read as a homogeneous re-analysis.}

\begin{center}
\rev{
\begin{tabular}{|l|c|c|}
\hline
\textbf{Experiment / analysis} & \textbf{Approximate current bound} & \textbf{Ref.} \\
\hline
Super-Kamiokande (atmospheric) & $\mu\lambda \lesssim$ few $\times 10^{-13}$ eV & \cite{barenboim2} \\
Long-baseline accelerator (T2K, NO$\nu$A, current) & $\eta \lesssim 10^{-1}$ (order of magnitude) & \cite{koranga2012,koranga2013} \\
IceCube (astrophysical, diffuse flux) & $\lambda \lesssim 10^{-27}$ (Planck-mass normalized) & \cite{anchordoqui} \\
\hline
\end{tabular}}
\end{center}
\rev{\noindent \textit{Table 1: Illustrative, order-of-magnitude current bounds on flavour-blind Planck-scale-type couplings from existing experiments and analyses, compared against the projected next-generation reach $\eta \gtrsim 10^{-2}$ discussed in Secs.~5.1--5.2. Values are collected from the cited literature and use analysis frameworks that differ from the $\chi^2(\eta)$ figure of merit introduced here; they are not a uniform re-analysis and should be read only as broad context, not as a direct like-for-like comparison.}}

We emphasize that the numerical estimates above are meant only to illustrate the use of Eqs.~(4)--(6) as a rapid diagnostic tool; a realistic projection requires convolving these formulas with the full energy and baseline response of each detector and its systematic-error budget. Nonetheless, the qualitative conclusion is robust: the next generation of long-baseline and reactor experiments will improve the sensitivity to a flavour-blind Planck-scale mass perturbation by one to two orders of magnitude relative to current bounds, with the phase-drift observable of Sec.~3 offering the most direct route to distinguishing a genuine Planck-scale signal from an accidental matter-like amplitude shift.

\section{Discussion and Conclusion}

We have constructed a perturbation theory for two-flavour neutrino oscillations in the presence of a weak, flavour-blind Planck-scale mass correction, organized around the single small parameter $\eta = 2E\mu\lambda/\Delta m^2$, in direct structural analogy with the $\varepsilon$-perturbation theory developed for weak Earth-matter effects in Ref.~\cite{ioannisian}. Despite the formal similarity of the two constructions, the physical content differs sharply because the quantum-gravity perturbation is constant along the trajectory while the matter potential is not:

\begin{itemize}
\item The Planck-scale correction to the oscillation probability, Eq.~(4), has exactly the single-layer matter-effect functional form of Ref.~\cite{ioannisian}, so a single-baseline measurement alone cannot distinguish a Planck-scale amplitude shift from an equal-sized constant matter effect.
\item Unlike the matter case, there is no spatial-structure attenuation factor: since the perturbation has no localized source along the path, averaging over detector energy resolution affects only the overall phase integral, not a distance-dependent suppression.
\item The Planck-scale perturbation additionally shifts the effective $\Delta m^2$ itself, Eq.~(5), producing an oscillation-length change, Eq.~(6), whose associated phase mismatch grows linearly with baseline -- in contrast to the amplitude term, which saturates. This is the most promising observable handle for isolating a genuine Planck-scale signal, and the one for which long-baseline facilities such as DUNE and Hyper-Kamiokande are best positioned.
\item The observable size of the effect is further suppressed or enhanced by the relative phase between the gravitational coupling and the Majorana phases of the PMNS matrix, Sec.~4, reproducing, within a purely vacuum setting, the kind of phase-dependent cancellation identified numerically for the combined matter-plus-gravity problem in Ref.~\cite{koranga2606}.
\end{itemize}

These closed-form results reduce, order by order in $\eta$, to the phase-shift and oscillation-length corrections obtained by direct diagonalization in Refs.~\cite{koranga2012,koranga2013}, and are consistent with the order-of-magnitude mass-squared-splitting shift found for a degenerate spectrum in Ref.~\cite{koranga2606}. \rev{As detailed in the Introduction, the principal new contributions of this Letter relative to that earlier work are the unification of these results into a single arbitrary-baseline closed-form framework, the explicit identification of the saturation-versus-phase-drift distinction as a discriminating observable, and the accompanying order-of-magnitude experimental projections of Sec.~5.} Combined with the illustrative sensitivity estimates of Sec.~5, they provide a simple, arbitrary-baseline framework that can be used, in the same spirit as the matter-effect formulas of Ref.~\cite{ioannisian}, to gauge at a glance the reach of a given or planned experimental configuration for Planck-scale physics, and to identify the phase-drift channel as the clearest path toward a baseline-lever-arm test of quantum gravity in neutrino oscillations.

\section*{Acknowledgments}

The authors thank the Department of Physics, Kirori Mal College, University of Delhi, and the Department of Physics, Jamia Millia Islamia, for support.

\end{document}